\documentstyle[12pt,epsf]{article}
\begin{document}
\setlength\textheight{8.75in}
\newcommand{\be}{\begin{equation}}
\newcommand{\ee}{\end{equation}}

\title{Bound monopoles in the presence of a dilaton}
\author{{\large Yves Brihaye} \\
{\small Facult\'e des Sciences, Universit\'e de Mons-Hainaut, }\\
{\small B-7000 Mons, Belgium }\\
{ } \\
   {\large Betti Hartmann}\\
{\small Department of Mathematical Sciences ,
University of Durham}\\
{\small Durham DH1 \ 3LE , United Kingdom}}

\date{\today}
%%%%%%\begin{titlepage}
\maketitle
\thispagestyle{empty}
\begin{abstract}
We study axially symmetric monopoles 
of both the SU(2) Yang-Mills-Higgs-Dilaton (YMHD) as well as of the
SU(2) Einstein-Yang-Mills-Higgs-Dilaton (EYMHD) system. We find that
equally to gravity, the presence of the dilaton field
can render an attractive phase. We also study the
influence of a massive dilaton on the attractive phase in
the YMHD system. 
\end{abstract}
\medskip
\medskip

\section{Introduction}
The Georgi-Glashow model with SU(2) gauge group 
constitutes the simplest non-abelian gauge field theory 
in which topological solitons exist~:
magnetic monopoles \cite{HP,poly}.
It consists of an SU(2) Yang-Mills theory coupled to a Higgs triplet
in the adjoint representation of the group and is spontaneously
broken by a Higgs potential (we refer to it as to the YMH model in 
the following). The solutions are characterized by their winding number
$n$, which arises due to topological arguments and is proportional 
to the magnetic charge of the configuration.
The solution with unit topological charge $n=1$
can be constructed with a spherically symmetric ansatz of the fields.
Since this
was found to be the unique spherically symmetric solutions \cite{bo},
the field configurations corresponding to  higher values of the topological
charge $n > 1$ (the multimonopoles)
involve at most axial symmetry and lead to systems
of partial differential equations \cite{rr,kkt}.
One feature of multimonopoles  is their instability~: for generic
values of the coupling constants of the theory the long-range repulsion
due to the gauge fields cannot be overcome by the short-range 
attraction due to the Higgs field. Only in the so-called
BPS (Bogomol'nyi-Prasad-Sommerfield) limit \cite{bo,ps} in which the Higgs 
field is massless
and therefore long-range, 
the two interactions exactly
compensate \cite{manton,weinberg}. The spatial components
of the stress-energy tensor were shown to vanish \cite{raif}
and thus  systems of non-interacting monopoles exist.

A few years ago, the YMH model was coupled to Einstein
gravity \cite{BFM} (resulting in a theory labelled EYMH)
and the spherically symmetric gravitating monopoles  with unit topological
charge were constructed. Also studied were the
corresponding non-abelian black holes solutions, which violate the "no-hair"
conjecture.
Quite recently \cite{hkk}, it was demonstrated that  
bound states of gravitating multimonopoles exist in the EYMH model.
Indeed, solving the equations for numerous values of 
the coupling constants, it was shown that two phases exist.
For small values of the Higgs coupling constant,
there exists a phase  for which the binding energy of the 
2-monopole and the 3-monopole is negative, leading to
classical solutions bounded by gravity.

On the other hand, it was pointed out \cite{FG},
that the coupling of the YMH system to a dilaton field (labelled YMHD)
renders regular classical solutions that 
share many properties with that of the EYMH model.

It is therefore natural to check if the coupling to a dilaton field can also
lead to systems of bound monopoles. The aim of this
paper is to study this question by analyzing the equations of the 
full EYMHD model incorporating both gravitation and a dilaton field.
Our numerical integration of the equations
 strongly indicate
that the analogy between the EYMH and the YMHD models
persist also for the multimonopole solutions.

In Sect.II we specify the model and its different components,
the axially symmetric ansatz and the relevant rescaling.
 The boundary conditions are presented in Sect. III. The numerical solutions
and their relevant features are
discussed in Sect. IV.
In particular, we study the effect
of the dilaton on both the solutions in flat and curved space
and also briefly discuss the implications of
a massive dilaton.

\section{SU(2) (Einstein-)Yang-Mills-Higgs-Dilaton 
 theory}

The action of the Yang-Mills-Higgs-Dilaton (YMHD) theory reads:
\begin{equation}
S=S_{M}=\int L_{M}\sqrt{-g}d^{4}x\ , 
\end{equation}
while for the Einstein-Yang-Mills-Higgs-Dilaton
(EYMHD) theory an additional term from the
gravity Lagrangian arises:
\begin{equation}
S=S_{G}+S_{M}=\int L_{G}\sqrt{-g}d^{4}x+ \int L_{M}\sqrt{-g}d^{4}x
\end{equation}
The gravity Lagrangian is given by :
\begin{equation}
L_{G}=\frac{1}{16\pi G}R
\end{equation}
where $G$ is Newton`s constant.\\
The matter Lagrangian is given in terms of the gauge field $A_{\mu}^a$, 
the dilaton field $\Psi$ and the Higgs field $\Phi^a$ ($a=1,2,3$):
\begin{equation}
L_{M}=-\frac{1}{4} e^{2\kappa\Psi}F_{\mu\nu}^{a}F^{\mu\nu,a}-
\frac{1}{2}\partial_{\mu}\Psi\partial^{\mu}\Psi-
\frac{1}{2}D_{\mu}\Phi^{a} D^{\mu}\Phi^{a}-e^{-2\kappa\Psi}V(\Phi^{a})-
\frac{1}{2}m^2\Psi^2
\label{lag}
\end{equation}
where $m$ denotes the mass of the dilaton field and
\begin{equation}
V(\Phi^{a})=\frac{\lambda}{4}(\Phi^{a}\Phi^{a}-v^2)^2
\end{equation}
The field strength tensor is given by:
\begin{equation}
F_{\mu\nu}^{a}=\partial_{\mu}A_{\nu}^{a}-\partial_{\nu}A_{\mu}^{a}+
e\varepsilon_{abc}A_{\mu}^{b}A_{\nu}^{c}
\end{equation}
and the covariant derivative of the in the adjoint representation
given Higgs field reads:
\begin{equation}
D_{\mu}\Phi^{a}=\partial_{\mu}\Phi^{a}+
e\varepsilon_{abc}A_{\mu}^{b}\Phi^{c}
\end{equation}
$e$ denotes the gauge field coupling, $\kappa$ the dilaton coupling, $\lambda$ the Higgs field coupling
and $v$ the vacuum expectation value of the Higgs field.\\

\section{Axially symmetric Ansatz}
For the metric, the axially symmetric Ansatz in isotropic
coordinates reads:
\begin{equation}
ds^2=
  - f dt^2 +  \frac{m}{f} \left( d r^2+ r^2d\theta^2 \right)
           +  \frac{l}{f} r^2\sin^2\theta d\varphi^2
\ , \label{metric} \end{equation}
where $f$, $m$ and $l$ are functions of $r$ and $\theta$ only.
In the special case of the YMHD system $m(r,\theta)
=l(r,\theta)=f(r,\theta)=1$.\\
The ansatz for the purely magnetic gauge field is 
\begin{eqnarray}
A_\mu dx^\mu =\frac{1}{2} A_\mu^a \tau^a dx^\mu &=& 
\frac{1}{2er} [ \tau^n_\phi 
 ( H_1 dr + (1-H_2) r d\theta ) \nonumber  \\
 &-& n ( \tau^n_r H_3 + \tau^n_\theta (1-H_4))
  r \sin \theta d\varphi ]
\ , \label{gf1}
\end{eqnarray}
and for the Higgs field the ansatz reads 
\begin{equation}
\Phi= \Phi^a\tau^a = \left(\Phi_1 \tau_r^{n}+\Phi_2 \tau_\theta^{n}\right)
\  \end{equation}
where the matter field functions $H_1$, $H_2$, $H_3$, $H_4$, $\Phi_1$ and
$\Phi_2$ depend only on $r$ and $\theta$.
The symbols $\tau^n_r$, $\tau^n_\theta$ and $\tau^n_\phi$
denote the dot products of the cartesian vector
of Pauli matrices, $\vec \tau = ( \tau^1, \tau^2, \tau^3) $,
with the spatial unit vectors
\begin{eqnarray}
\vec e_r^{\, n}      &=& 
(\sin \theta \cos n \varphi, \sin \theta \sin n \varphi, \cos \theta)
\ , \nonumber \\
\vec e_\theta^{\, n} &=& 
(\cos \theta \cos n \varphi, \cos \theta \sin n \varphi,-\sin \theta)
\ , \nonumber \\
\vec e_\phi^{\, n}   &=& (-\sin n \varphi, \cos n \varphi,0) 
\ , \label{rtp} \end{eqnarray}
respectively. Here, the topological charge $n$ enters the Ansatz for the fields.\\
The dilaton field $\Psi$ is a scalar field depending on $r,\theta$:
\begin{equation}
\Psi=\Psi(r,\theta)
\end{equation}
\subsection{Rescaling}
With the introduction of the dimensionless radial coordinate $x$ and
rescaling of the Higgs field, the dilaton field and the dilaton mass, respectively:
\begin{equation}
x\equiv rev\ , \ \phi=\frac{\Phi}{v}\ , \ \psi=\frac{\Psi}{v}\ , \
M_{dil}=\frac{m}{ev}\ ,
\end{equation}
the set of partial differential equations
depends only on three fundamental coupling constants:
\begin{equation}
\alpha=\sqrt{4\pi G}v,\ \ \beta=\frac{\sqrt{\lambda}}{e},\ \ \gamma=v\kappa
\end{equation}
where $\alpha=0$ in the YMHD system.
\subsection{Mass of the solution}
In the case of a massive dilaton ($M_{dil}\neq 0$),
the mass of the solution $\mu$ can be obtained from
integrating the Lagrangian density (\ref{lag}).
For $M_{dil}=0$, however, simple relations between the mass of the solution
and the derivative of the corresponding function at infinity exist.
In the YMHD system the mass 
is given in terms of the derivative of the dilaton field at infinity
\cite{FG}
\begin{equation}
\mu=\frac{1}{\gamma }\lim_{x\to\infty} x^2\partial_x\psi
\label{massdil}
\end{equation}
while in the EYMHD system it is given in terms of the derivative of the
metric function $f$ at infinity
\begin{equation}
\mu=\frac{1}{2\alpha^2 }\lim_{x\to\infty} x^2\partial_x f
\end{equation}
The mass $\mu_{ab}$ of the corresponding abelian solutions is given
in the EYMHD system by:
\begin{equation}
\mu_{ab}=(\alpha^2+\gamma^2)^{-1/2}
\end{equation}
with $\alpha=0$ in the limit of the YMHD system.

\section{Boundary conditions}
We look for regular, static, finite energy solutions that are asymptotically flat. 
The requirement of regularity leads to
the following boundary conditions at the origin:
\begin{equation}
\partial_{x}f(0,\theta)=\partial_{x}l(0,\theta)=
\partial_{x}m(0,\theta)=0,\ \ \partial_{x}\psi(0,\theta)=0
\end{equation}
\begin{equation}
H_i(0,\theta)=0,\ i=1,3 ,\ \ H_i(0,\theta)=1,\ i=2,4,\ \
\phi_i(0,\theta)=0,\ i=1,2 
\end{equation}
At infinity, the requirement for finite energy and asymptotically
flat solutions leads to the boundary conditions:
\begin{equation}
f(\infty,\theta)=l(\infty,\theta)=m(\infty,\theta)=1,\ \ \psi(\infty,\theta)=0
\end{equation}
\begin{equation}
H_i(\infty,\theta)=0,\ i=1,2,3,4 ,\ \ \phi_1(\infty,\theta)=1,\ \
\phi_2(\infty,\theta)=0 
\end{equation}
In addition,
boundary conditions on the symmetry axes (the $\rho$- and
$z$-axes) have to be fulfilled.
On both axes:
\begin{equation} 
H_1=H_3=\phi_2=0
\end{equation}
and
\begin{equation}
\partial_\theta f=\partial_\theta m=\partial_\theta l 
=\partial_\theta H_2=\partial_\theta H_4=
\partial_\theta \phi_1=\partial_\theta \psi=0
\end{equation}
\section{Numerical results}
Subject to the above boundary conditions, we have solved the system of partial differential
equations numerically.\
\subsection{Monopoles in YMHD theory}
It was noted recently \cite{hkk} that in a certain parameter
range of the coupling constants an attractive phase exists in the
EYMH system. Inspired
by the observation that the monopoles
in the YMHD system share many features with the monopoles in the EYMH system
\cite{FG},
we first studied the (multi)monopoles of the YMHD system in the limit of vanishing
dilaton mass.
We find that in the BPS limit ($\beta=0$)
there exists an attractive phase for all
values of $\gamma > 0$.
This is in close analogy to the EYMH system, where attraction
between the BPS monopoles exists for all $\alpha > 0$.
Indeed, the plot of the energy per winding number over $\gamma$ in the YMHD
system looks similar than Fig.~3 of \cite{hkk} when $\alpha$ is interchanged
with $\gamma$ and "Reissner-Nordstr\"om ($RN)$" is interchanged
with "Einstein-Maxwell-Dilaton ($EMD$)". Moreover, we find that
when comparing the quantity
\begin{equation}
\Delta E=\frac{E(n=1)}{1} - \frac{E(n=2)}{2}
\end{equation}
for the monopoles in the EYMH system for a specific value of $\tilde{\alpha}$ with
that of the monopoles in the YMHD system for a value of
$\gamma=\tilde{\alpha}$, the two values equal each other (at least within our
numerical accurancy). 

We were also interested in the implications of a massive dilaton. The massive
dilaton was previously considered only for the spherically symmetric solutions
\cite{FG2}. We studied the influence of the dilaton mass $M_{dil}$
on the attractive phase. Since now a mass is involved, the dilaton field
decays exponentially - contrasted to a power law decay in the
massless case - and the relation (\ref{massdil}) between
the derivative of the dilaton field at infinity and the mass of the 
solution is no longer valid.
Our numerical results are shown in Fig.~1, where we present the difference
between the mass (per winding number) of the $n=1$ solution and the
mass per winding number
of the $n=2$ solution $\Delta E= \frac{E(n=1)}{1}-\frac{E(n=2)}{2}$.
Clearly, the attraction is lost for $M_{dil} > \hat{M}_{dil}(\gamma)$.
We find that $\hat{M}_{dil}(\gamma=0.5)\approx 0.040$ and $\hat{M}_{dil}(\gamma=1.0)
\approx 0.028$, respectively.
Our numerical results
suggest further that for $M_{dil}\rightarrow\infty$ the value $\Delta E$ turns 
to zero indicating that
the monopoles are non-interacting in this limit. This is demonstrated in the following
table for $\gamma=1.0$~:

\begin{center}
\begin{tabular}{|l|l|}
\hline
$M_{dil}$ & $\Delta E$\\
\hline
$0. $ & $0.00737$\\
$0.01 $ & $0.00372$\\
$0.1 $ & $-0.01390$\\
$1. $ & $-0.01380$\\
$10. $ & $-0.00023$\\
\hline
\end{tabular}
\end{center}
This result can be understood
considering that for $M_{dil}\rightarrow\infty$ the 
dilaton function $\psi(x,\theta)$ has to turn to zero on the full interval
$x\epsilon$ $[0:\infty[$ for all $\theta$. Thus for the case
studied here ($\beta=0$), the BPS limit of the YMH system 
is recovered for $M_{dil}\rightarrow\infty$. Our numerical results strongly
support this interpretation. We find that with increasing $M_{dil}$
the dilaton field tends more and more to the trivial solution $\psi(x,\theta)=0$
and that the mass tends to one, which (in our rescaled variables) is just the mass
of the BPS solution in the YMH system.

\subsection{Monopoles in EYMHD theory}
Here, we only considered the case of $M_{dil}=0$.
We first studied the influence of the dilaton field on the attraction between
like monopoles in the limit of vanishing Higgs coupling (BPS limit). In Fig.~2,
we show the difference between the mass per winding number of the $n=1$ and the
$n=2$ 
solution $\Delta E= \frac{E(n=1)}{1}-\frac{E(n=2)}{2}$ for a fixed
$\alpha$ and varying $\gamma$. 
For $\alpha=0$ the limit $\gamma=0$ represents the BPS limit
of the YMH theory. The monopoles are non-interacting for $\beta=0$ and therefore 
the energy per winding number is equal for all (multi)monopole solutions
of different topological sectors. Since the YMHD-monopoles in the BPS limit reside
in an attractive phase for all $\gamma\neq 0$, $\Delta E$ should be positive, which
indeed, is demonstrated in Fig.~2.
For $\alpha=0.5$ the attraction between like monopoles in the EYMHD
system is  bigger than
in the pure EYMH system ($\gamma=0$) for all $\gamma\neq 0$.
The curve reaches a maximum of the difference at some value $\gamma$
and from there, the difference gets smaller. This can be understood from the fact that
for rising $\gamma$, the solutions tend to the EMD solutions which have mass per winding
number equal for all $n$. The curve for $\alpha=1.0$ shows the same behaviour apart from
the fact, that now for bigger $\gamma$ the attraction gets smaller than in the
$\gamma=0$ case. This is due to the fact, that in the pure EYMH system, the attraction 
has nearly reached its maximum at $\alpha=1.0$ and that
now inclusion of the dilaton field very soon makes the solution tend to 
a EMD solution.  

To study the influence of the dilaton on the monopole solutions for $\beta\neq 0$,
we followed \cite{hkk} and determined $\gamma_{eq}(\beta)$. This is - for a fixed
$\beta$ - the value of $\gamma$ for which the mass 
of the $n=1$ solution
is equal to the mass per winding number of the $n=2$ solution. For 
$\gamma < \gamma_{eq}$,
the mass per winding number of the $n=2$ is bigger than the mass of the $n=1$ solution
which implies that the monopoles are repelling, while for 
$\gamma > \gamma_{eq}$ it is smaller leading to an attractive phase.
Because globally regular solutions exist only for
$\gamma\leq\gamma_{max}^{n}(\beta)$ \cite{bhk}, the attractive phase 
is limited in parameter space by the $\gamma^{n=1}_{max}$ curve. (Since
$\gamma^{n=2}_{max}(\beta) > \gamma^{n=1}_{max}(\beta)$ for the values of
$\beta$ for which the attractive phase exists, the masses of the 
$n=1$ and $n=2$ solution can only be compared for
$\gamma\leq\gamma_{max}^{n=1}$.) For $\beta=\hat{\beta}$, the two curves meet
and no attractive phase is possible for $\beta > \hat{\beta}$.
In the EYMH system it was found that $\hat{\beta}\approx 0.21$ \cite{hkk}.
In Fig.~3, the values of $\gamma_{eq}$ and $\gamma_{max}^{n=1}$
are shown for three different values of $\alpha$. $\alpha=0$ represents the
YMHD system and  the $\gamma_{max}^{n=1}$- and $\gamma_{eq}$- curves look similar
than the $\alpha_{max}^{n=1}$- and $\alpha_{eq}$- curves of \cite{hkk}.
This again underlines the close analogy of the EYMH system and the YMHD system.

Comparing the three curves,
we find that the values of both
$\gamma_{max}^{n=1}$ and $\gamma_{eq}$ drop to smaller
values of $\gamma$ for fixed $\beta$ and increasing $\alpha$.
This results in the fact that the value of $\hat{\beta}$ seems to be 
independent on $\alpha$. For $\alpha_{1}$ the attractive 
phase is thus obtained for smaller values of $\gamma$ then in the 
$\alpha_{2}$ case, if $\alpha_{1} > \alpha_{2}$. It does not seem to exceed $\beta > 0.21$ for any value of
$\alpha$ though.
\section{Summary and concluding remarks}
We have studied axially symmetric dilatonic monopoles in flat and curved space.
In the limit of vanishing gravitational coupling and vanishing
dilaton mass, we find that
the presence of the dilaton field can render an attractive phase similar
to gravity. The close analogy between the EYMH and the YMHD system observed
in \cite{FG} thus persists for the multimonopoles.
When the dilaton field is massive, the attraction between the 
monopoles in the BPS limit of the YMHD system is lost 
for $M_{dil} > \hat{M}_{dil}(\gamma)$ and
the monopoles are repelling. For $M_{dil}\rightarrow\infty$, the dilaton function
has to turn to the trivial solution (to fulfill the requirement
of finite energy). The dilaton decouples
from the field equations and the pure YMH system, in which
the BPS monopoles are known to be non-interacting, is left.

When in the BPS limit both gravitation and the (massless) dilaton are 
coupled to the monopoles,
the value of $\Delta E=\frac{E(n=1)}{1}-\frac{E(n=2)}{2}$ - indicating the
strength of attraction - first increases from its value
in the EYMH system with increasing $\gamma$. $\Delta E$ reaches a maximum at
a value of $\gamma$ depending on $\alpha$ and from there decreases.       

In the non-BPS limit, the (massless) dilaton field is able to
overcome the long-range repulsion of the
gauge fields in a similar way than gravity. We find that the attractive phase
is limited in parameter space and that the value of $\beta$ for which
the attractive phase is lost is independent on $\alpha$.

While the $n=1$ monopole is stable due to the preservation of the
topological charge, the stability of
$n > 1$ monopoles is not obvious since it might be possible that they
decay
into singly charged monopoles thereby preserving the
total topological charge. We conjecture that the monpoles are stable as long
as they reside in the attractive phase.

We have studied axially symmetric monopoles for $n=2$ here. 
However, it was observed
that for $n\geq 3$ BPS monopoles with discrete symmetries exist \cite{HS}.
Since in the BPS limit the energy per winding number
is equal for all configurations (independent on the actual structure), it would
be interesting to construct these solutions in the (E)YMHD system.
Only then it could be decided, which configuration is the one
of lowest energy for a given topological sector. Moreover, computing the
energy per winding number of these solutions, the influence of gravity and the 
dilaton field, respectively, on monopoles with discrete symmetries could be
investigated. It has to be pointed out though, that up to now no
explicit Ansatz exists which would allow a numerical 
construction of these configurations. In the BPS limit, the monopoles
fulfill a first order differential equation, the Bogomol'nyi equation.
Since this equation is integrable, mathematical techniques such as twistor
methods \cite{sutcliffe} are available. In the non-BPS limit though,
the full system of second-order differential equations has to be solved.
This is a difficult numerical task and remains a challege for the future.
However, in analogy to the soliton solutions in the Skyrme model \cite{battye},
we conjecture that the actual minimal energy configurations of the system
studied here have rather discrete than axial symmetry for $n\geq 3$. 
Since our study of the binding energy of $n=3$ axial multimonopoles
shows that the domain of parameter space where
$n=3$ bound solutions exist varies only  little from the $n=2$ case, it is very 
likely that (-at least in the region of parameter space
we have studied-) the discrete symmetry solutions are even more 
binded than the axial ones we have constructed.  

{\bf Acknowledgements}
One of us (B.H.) wants to thank the Belgium F.N.R.S. for financial support.
We gratefully acknowledge discussions with J. Kunz and B. Kleihaus.

\small{
 
\newpage

\begin{figure}\centering\epsfysize=15cm
\mbox{\epsffile{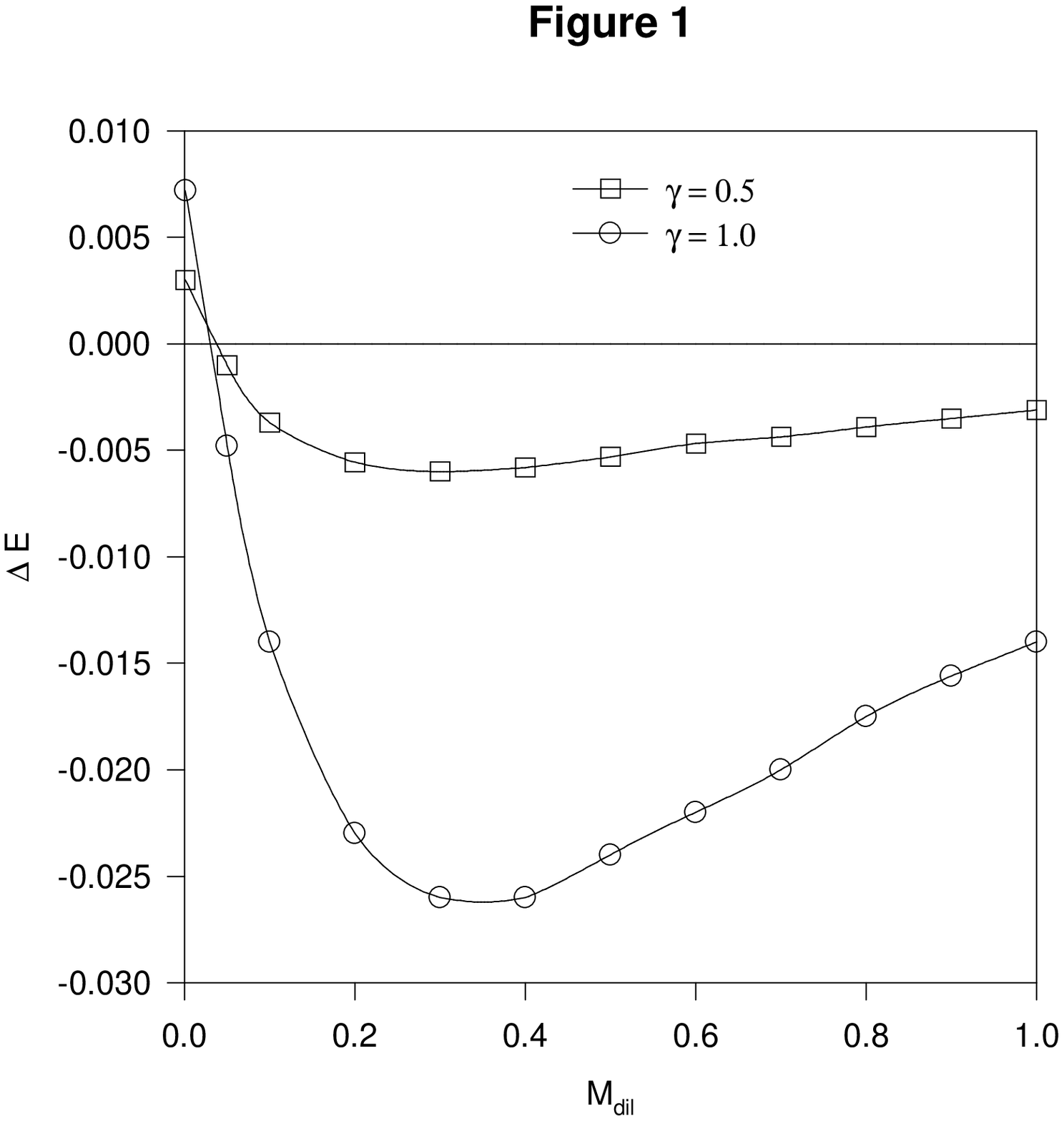}}
\caption{The quantity $\Delta E=
 \frac{E(n=1)}{1}-\frac{E(n=2)}{2}$ 
is shown as function of the dilaton mass $M_{dil}$
for two different values of $\gamma$ in the
YMHD system ($\alpha=0$).}
\end{figure}
\newpage
\begin{figure}\centering\epsfysize=15cm
\mbox{\epsffile{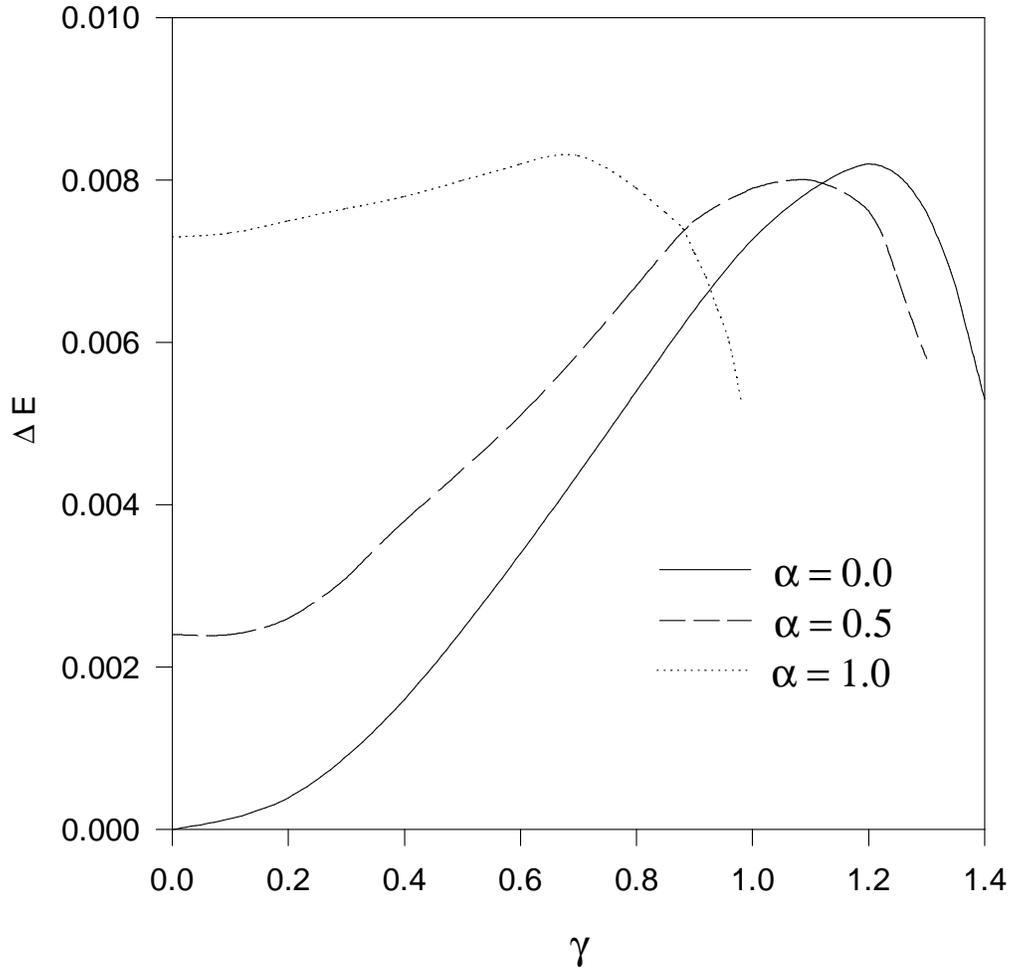}}
\caption{
The quantity $\Delta E= \frac{E(n=1)}{1}-\frac{E(n=2)}{2}$  is shown as function
of $\gamma$ for three different values of $\alpha$, including $\alpha=0.0$, which
represents the YHMD system.
}
\end{figure}
\newpage
\begin{figure}\centering\epsfysize=15cm
\mbox{\epsffile{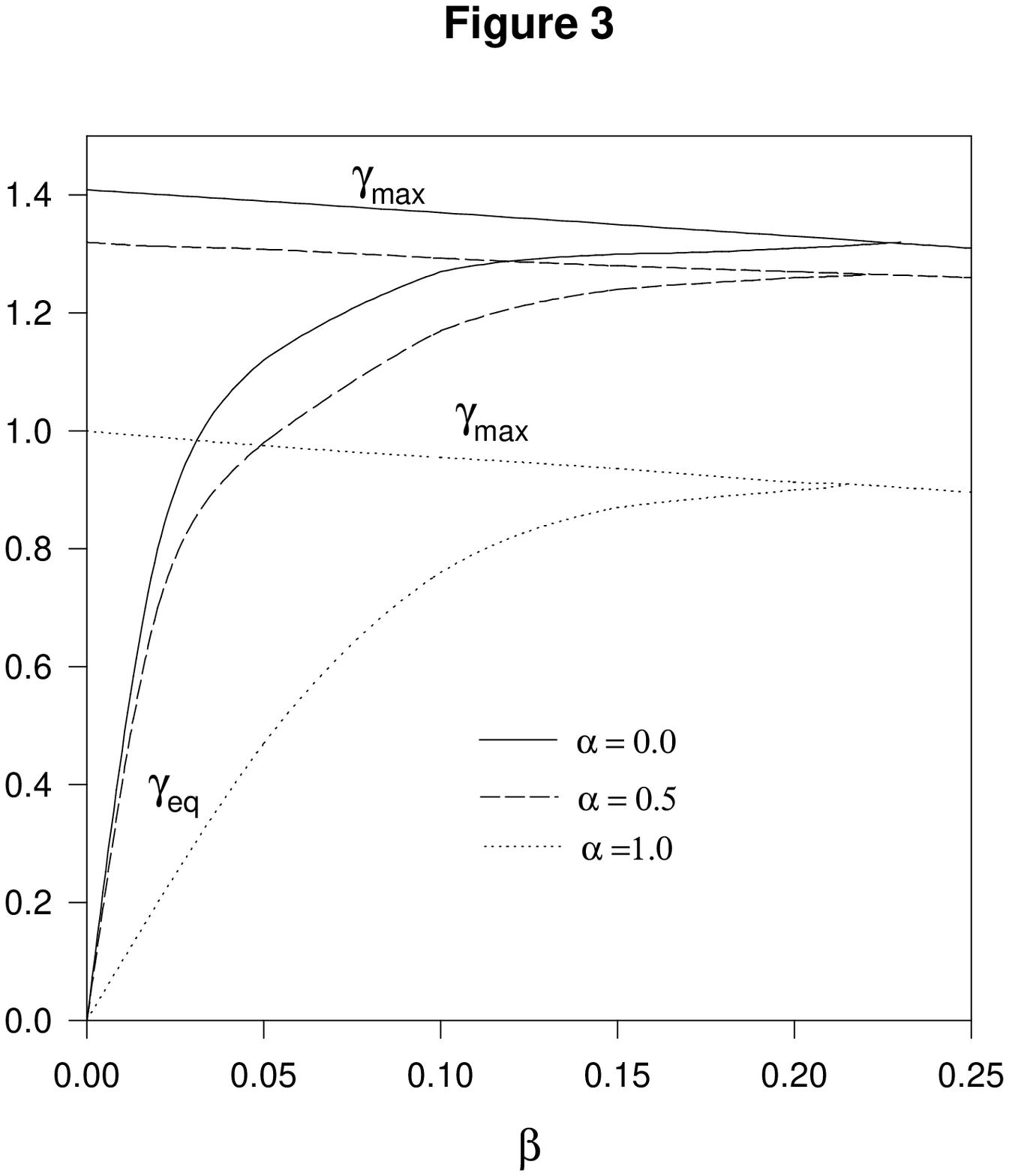}}
\caption{$\gamma_{eq}$ is shown as a function of $\beta$ 
for three different values of $\alpha$. Also shown is $\gamma_{max}^{n=1}$. 
The attractive phase exists for parameters values
above the $\gamma_{eq}$ curve and below the corresponding $\gamma_{max}^{n=1}$ 
line.
}

\end{figure}
\end{document}